\documentclass[prl,aps,amssymb,twocolumn,superscriptaddress,notitlepage,reprint,showpacs,nobalancelastpage]{revtex4-1}
\usepackage{amsmath}
\usepackage{amssymb}
\usepackage{amsthm}
\usepackage{amsfonts}
\usepackage{listings}
\usepackage{enumerate}
\usepackage{latexsym}
\usepackage{psfrag}
\usepackage{bm}
\usepackage{graphicx}
\usepackage[caption=false]{subfig}
\usepackage{blkarray}
\usepackage{array}
\usepackage{color}
\usepackage[normalem]{ulem}
\usepackage{hyperref}
\usepackage{dsfont}
\usepackage{ulem}

\usepackage{sidecap}

\def\bk{{\sf k}}

\def\bt{{\sf t}}

\def\bM{{\sf M}}

\def\bG{{\sf G}}

\def\bB{{{\sf B}}}

\def\bb{{{\sf b}}}

\newcommand{\beq}{\begin{equation}}
\newcommand{\eeq}{\end{equation}}
\newcommand{\beqarray}{\begin{eqnarray}}
\newcommand{\eeqarray}{\end{eqnarray}}

\begin{document}
\title{Partial lattice defects in higher order topological insulators}

\author{Raquel Queiroz}
\email{raquel.queiroz@weizmann.ac.il}
\affiliation{Department of Condensed Matter Physics,
Weizmann Institute of Science,
Rehovot 7610001, Israel}

\author{Ion Cosma Fulga}
\email{i.c.fulga@ifw-dresden.de}
\affiliation{Institute for Theoretical Solid State Physics, IFW Dresden, Helmholtzstr. 20, 01069 Dresden, Germany}

\author{Nurit Avraham}
\affiliation{Department of Condensed Matter Physics,
Weizmann Institute of Science,
Rehovot 7610001, Israel}

\author{Haim Beidenkopf}
\affiliation{Department of Condensed Matter Physics,
Weizmann Institute of Science,
Rehovot 7610001, Israel}

\author{Jennifer Cano}
\email{jennifer.cano@stonybrook.edu}
\affiliation{Princeton Center for Theoretical Science, Princeton University, Princeton, New Jersey 08544, USA}
\affiliation{Department of Physics and Astronomy, Stony Brook University, Stony Brook, New York 11974, USA}
\affiliation{Center for Computational Quantum Physics, Flatiron Institute, New York, New York 10010, USA}

\date{\today}
\begin{abstract}
Non-zero weak topological indices are thought to be a necessary condition to bind a single helical mode to lattice dislocations.
In this work we show that  higher-order topological insulators (HOTIs) can, in fact, host a single helical mode along screw or edge dislocations (including step edges) in the absence of weak topological indices. When this occurs, the helical mode is necessarily bound to a dislocation characterized by a fractional Burgers vector, macroscopically detected by the existence of a stacking fault. 
The robustness of a helical mode on a partial defect is demonstrated by an adiabatic transformation that restores translation symmetry in the stacking fault. 
We present two examples of HOTIs, one intrinsic and one extrinsic, that show helical modes at partial dislocations.
Since partial defects and stacking faults are commonplace in bulk crystals, the existence of such helical modes can measurably affect the expected conductivity in these materials. 
\end{abstract}
\maketitle

\paragraph{\bf Introduction}

Topological insulators (TI) with weak indices \cite{Fu2007,Moore2007,Roy2009} have the 
distinctive property of hosting single one-dimensional (1D) helical modes on line dislocations \cite{Ran2009,Teo2010defects,Teo2017review}. 
These topological modes can be regarded as more robust than the helical surface states of a weak topological insulator (WTI) because they do not require translation symmetry for protection \cite{Ran2009,Teo2010defects,Ringel2012}.
Thus, they are a valuable tool to identify and probe the physics of WTIs experimentally \cite{Tretiakov2010,Sbierski2016,Hamasaki2017}.
Beyond WTIs, the existence of protected gapless modes localized on topological defects generalizes to band insulators in other dimensions and symmetry classes \cite{Teo2010defects,Juricic2012,Asahi2012,Ruegg2013,Hughes2014,deJuan2014}, as well as to topological band insulators protected by crystal symmetry \cite{Teo2013,Shiozaki2014,Slager2014,Benalcazar2014,Bradlyn2017,Cano2018,Wieder2018}.

\textit{Partial} dislocations -- those whose Burgers vector is a fraction of a lattice translation -- fall outside of the topological classification: because partial dislocations are necessarily accompanied by a stacking fault plane, they are locally detectable arbitrarily far away from the dislocation line and, thus, do not constitute a topological defect.
However, as we show in the present manuscript, partial dislocations can host topologically protected gapless modes.
Since multiple partial dislocations can combine to form a full dislocation, consistency with the classification in Ref.~\onlinecite{Teo2010defects} provides conditions under which a partial dislocation can exhibit a gapless topological mode.

We find that the existence of topological modes on partial defects is intimately related to the recently predicted higher-order topological insulators (HOTIs) \cite{Benalcazar2017science,Schindler_HOTI,Song2017,Langbehn2017,Benalcazar2017prb,Geier2018,Trifunovic2018,Schindler_Bi,Khalaf2017,Khalaf2018,Ezawa2018phosphorene,Ezawa2018b,Matsugatani18,Imhof2017,Peterson2018,SerraGarcia2018,Noh2018,You2018}.
HOTIs of order $d$ in $D$ spatial dimensions are characterized by gapless topological modes on their $D-d$ dimensional edges. These gapless modes reside between $(D-d+1)$-dimensional surfaces that are gapped by mass terms of different sign. If the mass term on either side of the $(D-d)$-dimensional edge is forced to differ in sign because the corresponding $(D-d+1)$-dimensional surfaces are related by symmetry, then the HOTI is \textit{intrinsic}: its topological edge mode cannot be removed without closing the bulk gap or breaking crystal symmetry. 
In contrast, the gapless $(D-d)$-dimensional edge modes of \textit{extrinsic} HOTIs can be removed while preserving the bulk gap and crystal symmetry, by closing the surface gap \cite{Geier2018}. 
HOTIs, like WTIs and topological crystalline insulators \cite{Fu2011,Hsieh2012}, have a trivial bulk {\sout{and surface}} in the absence of crystal symmetries.

In this manuscript, we focus on $D\!=\!3,~d\!=\!2$.
We prove that a system whose partial dislocations host a gapless topological 1D mode must either have gapless surface states or realize an intrinsic or extrinsic HOTI for some surface termination.
Focusing on symmetry class AII \cite{Altland1997} -- although our results can be generalized -- we present two models, of an intrinsic and an extrinsic second order topological insulator, which have trivial weak indices, but which realize gapless helical modes on partial screw dislocations.
Both models can be smoothly deformed to a WTI by symmetrizing the Hamiltonian such that the unit cell is halved.
During this process, the bulk gap remain open, while the surface gap closes; thus, the deformation of the Hamiltonian is accompanied by an insulator-to-metal surface phase transition.
The existence of a partial Burgers vector in a HOTI that can be elevated to a full lattice vector in a WTI without closing the bulk gap is a sufficient condition to realize a helical mode on a partial dislocation.
More generally, we prove that a gapless helical mode on a partial lattice dislocation is an unambiguous signature of topology and that, in the reverse direction, for every HOTI with gapless helical modes on its hinges, there exists a lattice defect that exhibits a gapless helical mode.

The existence of helical modes bound to defects is a valuable probe to experimentally detect HOTIs in cases where the $(D-d)$-dimensional (hinge) modes are not visible.

\begin{figure}[t!]
    \centering
    \includegraphics[width=1.\columnwidth]{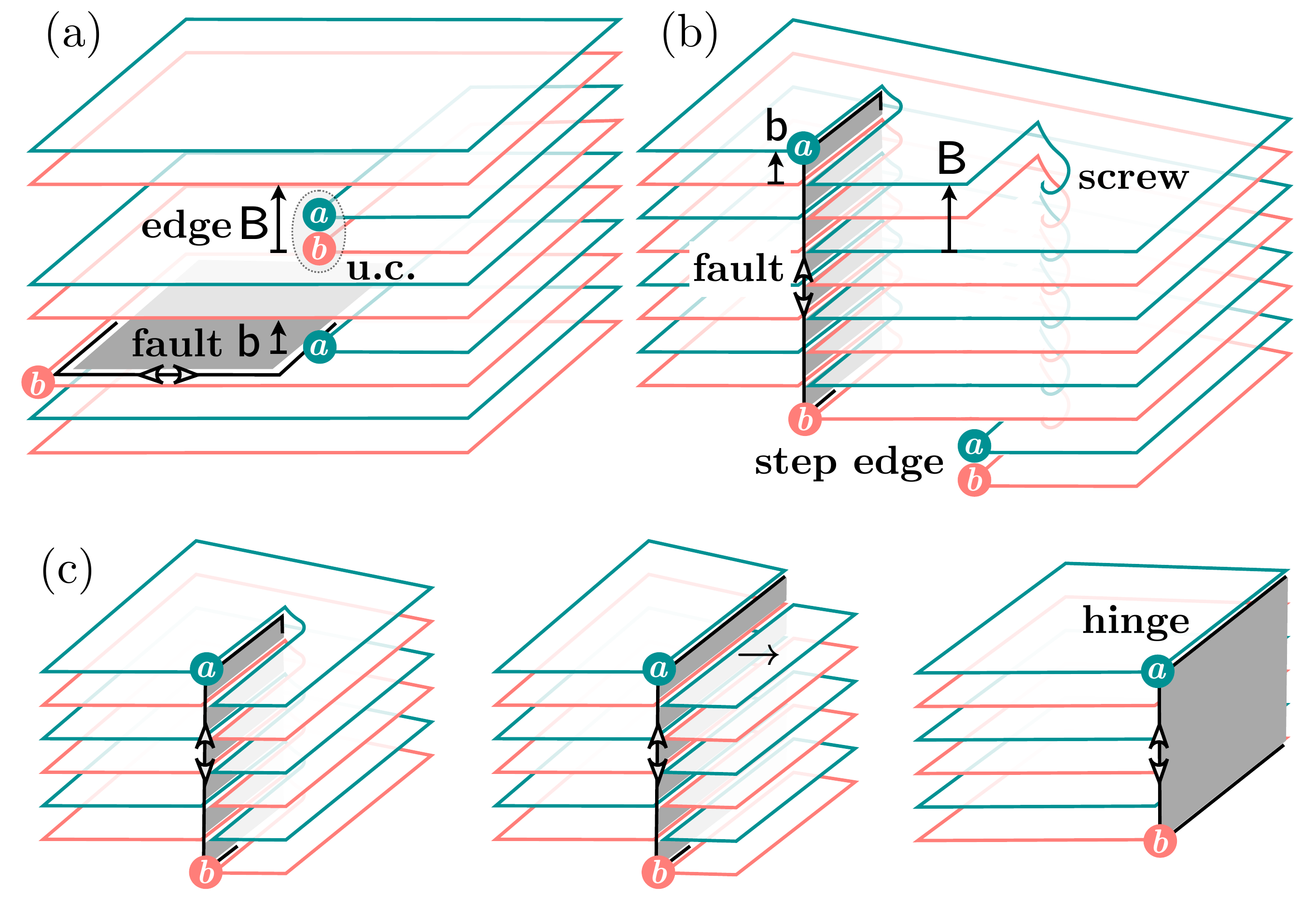}
    \caption{Full and partial dislocations with Burgers vectors $\bB$ and $\bb$, respectively; $a$ and $b$ indicate sublattice degrees of freedom in the unit cell. $\bB$ is a lattice translation, while $\bb$ is not. 
    The dislocations are either (a) edge dislocations or (b) screw dislocations.
    A full dislocation is locally invisible away from its core.
    In contrast, a partial dislocation is attached to a 2D stacking fault (gray plane). 
    This partial dislocation can host a helical mode, indicated by a black line and double-arrow.
    (c) Adiabatic deformation of the lattice, taking the stacking fault into the surface. The helical mode localized at a screw dislocation mode can be moved into the hinge.}
    \label{fig:helical_mode_scheme}
\end{figure}

\paragraph{\bf Partial screw and edge dislocations}
\label{sec:partialscrew}

A dislocation line is characterized by a lattice vector $\bB$, the Burgers vector. 
The dislocation breaks lattice translation symmetry, but away from the line, translation symmetry is restored, rendering the line dislocation locally invisible. 
Hence, a line dislocation is a \textit{topological} defect because it is only detectable by a non-local probe: a loop around the line dislocation can only be closed with an additional translation of $\bB$ relative to the same loop without the defect. 
When a dislocation terminates at a surface, it results in a step edge, depicted in Fig.~\ref{fig:helical_mode_scheme}.

Edge and screw dislocations are distinguished by whether $\bB$ is perpendicular or parallel to the defect, respectively. 
A general dislocation can be a combination.

In a crystalline system, topological defects are classified by how the Bloch Hamiltonian winds as it is transported around the defect \cite{Teo2010defects}. In three-dimensional time-reversal invariant systems with spin-orbit coupling (class AII), a line defect is classified by a $\mathbb{Z}_2$ invariant, which indicates whether it hosts a helical mode \cite{Teo2010defects}. 
The $\mathbb{Z}_2$ invariant corresponding to a dislocation described by $\bB$ is determined completely by the weak indices of the Hamiltonian \cite{Ran2009}: it is nontrivial if
\begin{equation}
\bB\cdot\bM_\nu=\pi\!\mod2\pi,
\label{eq:weakindex}
\end{equation}
where the time-reversal invariant momentum
$\bM_\nu=(\nu_1\bG_1+\nu_2\bG_2+\nu_3\bG_3)/2$ is determined by the weak topological indices $(\nu_1,\nu_2,\nu_3)$ and the reciprocal lattice vectors $\bG_i$ \cite{Fu2007,Moore2007,Roy2009}.

Here, we consider topological helical modes at dislocations that are characterized by a Burgers vector that is 
\emph{not} a lattice vector, which we denote by $\bb$.
Such a defect is referred to as a \textit{partial} dislocation \cite{dislocationbook}.
A partial dislocation is always bound to a stacking fault -- a 2D plane where translation symmetry is broken -- as shown in Fig.~\ref{fig:helical_mode_scheme}.
Thus, a partial dislocation is not a topological defect and, consequently, 
Eq.~\eqref{eq:weakindex} does not apply. 
In fact, we will show that a system with trivial weak indices can host a single gapless helical mode on a partial dislocation.

To gain insight into which partial dislocations can host helical modes, 
we derive a general constraint by combining multiple partials to form a full dislocation:
suppose $\bb$ characterizes a partial dislocation that hosts $h$ gapless helical modes, and define $n>1$ be the minimum integer such that $n\bb$ is a lattice vector.
(The case $n=2$ is depicted in Fig.~\ref{fig:helical_mode_scheme}.)
Then consider the full dislocation characterized by $n\bb$.
This dislocation hosts $nh\mod 2$ helical modes.
Comparison with Eq.~\eqref{eq:weakindex} requires 
\begin{equation}
n\bb\cdot \bM_\nu = nh\pi \mod 2\pi.
\label{eq:nb}
\end{equation}
There are four cases:
First, if $n$ is even and $n\bb\cdot \bM_\nu=0$, then any value of $h$ satisfies Eq.~\eqref{eq:nb}; in particular,
a system may have trivial weak indices and yet host gapless helical modes on partial defects. 
This is the case considered in the models that follow.
Second, if $n$ is even and $n\bb\cdot \bM_\nu=\pi$, then there is no $h\in \mathbb{Z}$ that satisfies Eq.~\eqref{eq:nb}; hence, the stacking fault that accompanies $\bb$ is must be gapless. 
Third, if $n$ is odd and $n\bb\cdot \bM_\nu = \pi$, then $h$ must be odd and hence the partial screw dislocation \textit{must} host a gapless helical mode if the stacking fault is gapped.
Finally, if $n$ is odd and $n\bb\cdot \bM_\nu =0$, then $h$ must be even: the partial defect \textit{cannot} host a single gapless helical mode. A look-up table summarizes these results in Supplementary Material (SM)  Sec.~A \cite{SM}.

\paragraph{\bf Connection to HOTIs}
We now argue that a system that realizes a helical mode on a partial dislocation is an (extrinsic or intrinsic) HOTI or has gapless surface states.
This follows because a step edge can be deformed into a ``hinge'' between two surfaces of a crystal, depicted in Fig.~\ref{fig:helical_mode_scheme}c and shown numerically in SM Sec.~B \cite{SM}.
Specifically, if a helical mode exists on a partial dislocation, then the same mode must exist on a partial step edge where the dislocation terminates on a surface.
The partial step edge can be turned into a hinge by adding or removing atoms until it reaches the edge of the crystal.
Because the topological protection of the helical mode only depends on time-reversal symmetry, it will be robust to this deformation provided it does not encounter another gapless mode, either on the hinge (in which case the material is already a HOTI) or on the side surface.
Thus, this construction yields a surface termination with a gapless helical mode; hence, it is a HOTI.
However, the existence of the helical mode on a partial dislocation does not depend on the surface termination; it is a bulk characteristic.

In the reverse direction, given a HOTI with gapped surfaces (which excludes a WTI) and gapless helical modes on its hinges, a defect with a gapless helical mode can be engineered by ``stacking'' in real space multiple copies so that an external hinge mode becomes an internal defect and translation is broken across the stacking plane. Depending on the spatial embedding of the degrees of freedom in the HOTI, this defect may be a partial dislocation; it cannot be a full dislocation (otherwise, it would be a WTI). 
If the Hamiltonian can be smoothly deformed to preserve translation symmetry
across the stacking plane, then the defect was necessarily a partial dislocation and the deformed system is a WTI with a helical mode on a full dislocation.

\paragraph{\bf Models}
\label{sec:models}

We present two models of 3D HOTIs, characterized by a gapped bulk and gapped surfaces, but gapless helical modes along one-dimensional edges.
They are both ``doubled'' models, constructed from two interpenetrating sublattices that separately realize a topological phase (either a weak or strong TI), but combined the system has trivial weak and strong indices.
Both models can be continuously deformed to a WTI without closing the bulk gap by turning off the inter-sublattice coupling, which halves the unit cell.
When the sublattices are coupled, they have trivial weak (and strong) indices, but nonetheless host gapless helical modes on certain partial screw dislocations.
In SM Sec. C, we prove that the helical modes are required by computing the nontrivial $\mathbb{Z}_2$ invariant of the stacking fault.

\begin{figure}[tb]
\centering
\includegraphics[width=1\columnwidth]{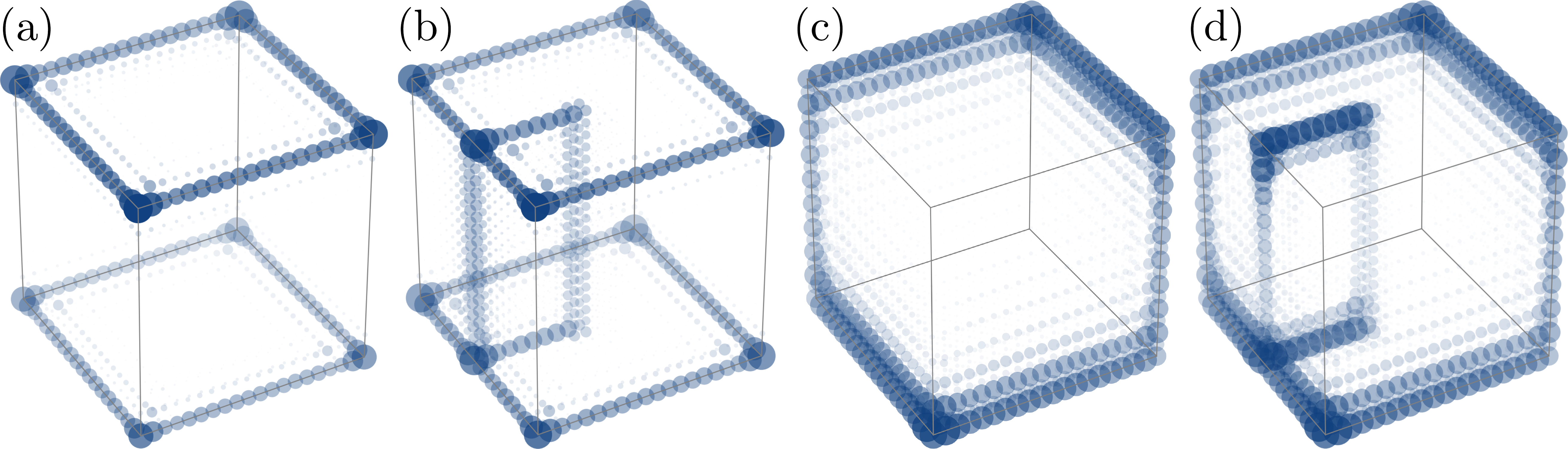}
\caption{HOTIs with screw dislocations. (a) Hinge modes of the extrinsic HOTI (Eqs.~\eqref{eq:2DTI} and \eqref{eq:dimerWTI}). 
(b) The extrinsic HOTI with a single partial screw dislocation. 
(c) Hinge modes of the intrinsic inversion-protected HOTI (Eqs.~\eqref{eq:DSTI} and \eqref{eq:DSTIdelta}).
(d) The intrinsic HOTI with a single partial screw dislocation. 
In all panels, the real-space probability distribution is averaged over the eight states closest to $E=0$. Larger circles and darker colors correspond to larger probability densities. 
}
\label{fig:HOTI}
\end{figure}

\paragraph{Extrinsic HOTI}

We start with a WTI constructed from 2D quantum spin Hall (QSH) layers stacked evenly with spacing $\hat{z}$. 
We add a perturbation that alternates the coupling between adjacent layers, doubling the unit cell without closing the bulk gap; this is a generalization of the Su-Schrieffer-Heeger chain \cite{Su1979}. 
Terminating the system between strongly coupled layers will reveal 1D helical modes along its hinges, while terminating between weakly coupled layers yields gapped hinges.
This model comprises an \emph{extrinsic} HOTI \cite{Geier2018}, since the presence/absence of helical hinge modes depends on the surface termination. Consequently, the extrinsic HOTI is not symmetry indicated \cite{Po2017,Bradlyn2017}: the quantum numbers associated to wavefunctions at high-symmetry points in the Brillouin zone do not yield a bulk topological invariant.

We consider the first quantized Hamiltonian $H(\bk)=H_{0}(\bk)\mu_0+H_{\delta}(\bk)$
where $\bk$ labels the crystal momentum, $\sigma_z$ and $\tau_z$ the spin and orbital degrees of freedom, and $\mu_z$ a sublattice index labelling the two inequivalent sites in the dimerized unit cell. We define
\begin{align}
H_{0}(\bk)=M(\bk)\tau_z-A(\sin k_x\sigma_x-\sin k_y \sigma_y)\tau_x,
\label{eq:2DTI}
\end{align} 
with
$M(\bk) = M-B(4-2\cos k_x - 2\cos k_y)$ and $0<M<4B$, which
describes a 2D TI in each layer, and
\begin{equation}
H_{\delta}(k_z)\!=\![ (t-\delta)\mu_y-(t+\delta)(\cos k_z\mu_y-\sin k_z\mu_x)]\sigma_z\tau_x,
\label{eq:dimerWTI}
\end{equation}
which couples the layers in a dimerized fashion.
When $\delta=0$, $H$ describes a WTI with indices $(0;001)$ and a lattice translation of $\hat{z}$. 
Away from this fine-tuned point, when $\delta \neq 0$, but $\delta$ is much smaller than the bulk gap, the system remains adiabatically connected to a WTI,
but its unit cell is doubled and, consequently, its $\mathbb{Z}_2$ indices are trivial, $(0;000)$. 
The new cell causes the Brillouin zone to fold; hence, when $\delta\neq 0$, the gapless surfaces of the WTI (at $\delta = 0$) become gapped.
Furthermore, when $\delta \neq 0$, the system can be terminated such that two gapless helical modes reside along its top and bottom edges (Fig.~\ref{fig:HOTI}a); thus, it is an extrinsic HOTI.

When $\delta=0$, a screw dislocation that connects two adjacent layers ($\bB=\hat{z}$) is a full dislocation that hosts a helical mode according to Eq.~\eqref{eq:weakindex}.
When $\delta \neq 0$, the helical mode remains (its topological protection does not rely on translation symmetry), but since $\hat{z}$ is not a lattice vector, the dislocation is partial.
Hence, this model has trivial weak indices and a helical mode along a partial dislocation.

Using the Kwant code \cite{Groth2014}, we have numerically implemented this model on a finite size sample of $20^3$ sites, setting $M=2$, $A=B=1$, $t=0.3$, and $\delta=0.2$.  Fig.~\ref{fig:HOTI}a shows the helical modes on the top and bottom surfaces. Fig.~\ref{fig:HOTI}b shows the partial screw dislocation with $\bb = \hat{z}$, confirming that the gapless helical mode is bound to the dislocation core as well as to step edges that emanate from it. (See SM Sec.~D for band structure.)
Our numerical simulation shows a surface termination realizing the extrinsic HOTI phase.
However, since the helical mode is a bulk feature, it will reside on the screw dislocation regardless of the surface termination.

\paragraph{Intrinsic HOTI}
Our second model consists of two coupled strong 3D TIs, one on each of the two sublattices (labelled $a$ and $b$).
Each 3D TI is described by:
\begin{align}
H_{0}(\bk)\!=\!M(\bk)\tau_z + A(\sin k_x\sigma_z  \tau_x - \sin k_y\tau_y + \sin k_z\sigma_x  \tau_x),
\label{eq:DSTI}
\end{align}
where $M(\bk) = M-B(6-2\cos k_x -2\cos k_y-2 \cos k_z)$.
$H_{0}(\bk)$ obeys time-reversal symmetry, $\mathcal{T} = i\sigma_y K$, where $K$ is complex conjugation, and inversion symmetry, $\mathcal{P}=\tau_z$.
We consider the regime $0<M<4B$, where there is one occupied Kramers pair (at $\Gamma$) with negative inversion eigenvalues.

We now introduce a sublattice degree of freedom, indexed by $\mu_z$.
The Hamiltonian $H_{0}(\bk)\mu_0$ describes two (uncoupled) 3D TIs.
This model was introduced as the ``double strong TI'' (DSTI) \cite{Khalaf2017}, without specifying the spatial embedding of the sublattices.
Here, we offset the $b$ sublattice half a unit cell in the $\hat{z}$ direction.
This shift preserves the inversion center about the origin.
It also introduces a translation symmetry by $\hat{z}/2$, denoted $\bt_{\hat z/2}$, which exchanges the two sublattices.
The extra translation symmetry causes the Brillouin zone to unfold so that the two band inversions are now located at $\Gamma$ and $Z\equiv (0,0,\pi)$.
Consequently, $H_{0}(\bk)\mu_0$ describes a WTI with indices $(0;001)$.

We add a perturbation that breaks $\bt_{\hat z/2}$ down to $\bt_{\hat z}$ and gaps all surfaces:
\begin{equation}
H_{\delta}(\bk) = m\sin \frac{k_z}{2}\sigma_y\tau_x \mu_x+\delta\cos \frac{k_z}{2}(\sigma_x  \tau_z +\sigma_y  \tau_0)  \mu_y.
\label{eq:DSTIdelta}
\end{equation} 
The first term preserves $\bt_{\hat z/2}$ symmetry, but gaps the surface Dirac cones (one from each sublattice) on the $\hat{z}$-normal surface.
The second term breaks $\bt_{\hat z/2}$ symmetry and thus gaps the Dirac cones on the $\hat{x}$- and $\hat{y}$-normal surfaces; in an electronic system, this can emerge from a charge density wave or a Jahn-Teller distortion. 
Thus, $H(\bk)=H_{0}(\bk)\mu_0+H_\delta(\bk)$ has a gapped bulk and surfaces.
Its $\mathbb{Z}_2$ indices are trivial, $(0;000)$.
However, the inversion eigenvalues of the occupied bands yield a nontrivial HOTI index of $2\text{ mod }4$ \cite{Khalaf2017}.
This model realizes an \emph{intrinsic} HOTI: a finite sample realizes a single helical mode that cannot be removed without breaking inversion symmetry.
(If inversion symmetry is broken, the helical mode will persist provided all bulk and surface gaps remain open.)

We now insert partial screw dislocation with $\bb=\hat{z}/2$.
When $m=\delta=0$ and $\bt_{\hat z/2}$ symmetry is preserved, this is a full dislocation that hosts a single helical mode according to Eq.~(\ref{eq:weakindex}).
When $m,\delta\neq 0$, the dislocation is partial, but as long as $H_\delta(\bk)$ does not close the bulk gap, the helical mode must survive.
Thus, we have provided an example of an intrinsic HOTI in which a partial screw dislocation hosts a gapless helical mode, while the bulk has trivial $\mathbb{Z}_2$ indices.

We numerically implemented this model on a finite size sample of $20\times20\times19$ sites, with $M=2$, $A=B=1$, $m=2$, and $\delta=0.5$. Fig.~\ref{fig:HOTI}c shows the single helical mode that traverses an inversion-symmetric path across the hinges. Fig.~\ref{fig:HOTI}d shows the partial screw dislocation with $\bb=\hat{z}/2$, which hosts a gapless helical mode along the dislocation as well as on the step edges on the top and bottom surfaces.
(See SM Sec.~D for band structure.)

\paragraph{Edge dislocations} Because screw and edge dislocations can be deformed into each other, edge dislocations in HOTIs can also realize gapless helical modes, which we demonstrate in SM Sec E.

\paragraph{\bf Measurement}
Partial lattice defects are ubiquitous in crystals \cite{dislocationbook}.
Their presence can be detected by a surface step edge, whose height reveals whether it is partial or full.
The density of states on the step edge can be measured via scanning tunneling spectroscopy (STS), as has been demonstrated for several topological materials \cite{Drozdov2014,Pauly2015,Sessi2016,Wu2016,Li2016,Fedotov2017,Peng2017,Jia2017,Schindler_Bi,Iaia2018,Yam2018}.

Recently, $1T'$-MoTe$_2$ and $1T'$-WTe$_2$ were predicted to be intrinsic HOTIs \cite{Wang2018}.
The latter is unstable in bulk form; however, STS reveals topological edge modes on single-layer $1T'$-WTe$_2$ \cite{Peng2017,Jia2017}.
Since the bulk unit cell contains two such layers, a single-layer step edge is a partial dislocation; thus, the measurement is consistent with our analysis.
While Bismuth \cite{Schindler_Bi} and SnTe \cite{Schindler_HOTI} have also been predicted to be intrinsic HOTIs and
zero bias peaks have been measured on full step edges in both materials \cite{Drozdov2014,Schindler_Bi,Sessi2016}, they fall outside of the scope of our work because neither can be adiabatically connected to a WTI.

We propose a candidate \textit{extrinsic} HOTI, Bi$_{13}$Pt$_{3}$I$_{7}$, which consists of dimerized 2D TI layers \cite{Pauly2015}, similar to the extrinsic HOTI model considered in this manuscript.
Consistent with trivial weak topological indices, STS measurements revealed that full step edges are gapped \cite{Pauly2015}; partial step edges were not studied.
We predict that a partial step edge in Bi$_{13}$Pt$_{3}$I$_{7}$ hosts a gapless helical mode.

Helical modes on dislocations can also be measured by their effect on conductivity \cite{Sbierski2016,Hamasaki2017}.

\paragraph{\bf Discussion}

Our work is a first step in the analysis of partial lattice defects in topological phases.
Partial dislocations fall outside the scope of Eq.~(\ref{eq:weakindex}). 
Nonetheless, we have shown that when a partial Burgers vector can be continuously deformed to a lattice vector by restoring a translation symmetry, the partial dislocation becomes a full dislocation that can host a gapless helical mode subject to the constraint of Eq.~(\ref{eq:weakindex}). 
We have shown that helical modes on partial dislocation lines can be used to experimentally detect HOTIs since they provide a sufficient condition for higher order topology.
Furthermore, in every HOTI it is possible to construct a defect with a gapless helical mode.

There are many possible future directions. 
The analysis will generalize beyond class AII.
Furthermore, a bulk invariant that captures both the intrinsic and extrinsic HOTI models remains an open question; recent progress has been made in Ref.~\cite{Khalaf2019}.
We show in SM C that the low-energy theory of the stacking fault can be regarded as an embedded 2D topological insulator.
Hence, the entanglement diagnosis in Ref.~\onlinecite{Tuegel2018} could potentially identify partial lattice defects that host topological modes in more general models.

\paragraph{\bf Acknowledgements} We thank Binghai Yan, Abhay Kumar-Nayak, and Jonathan Reiner for helpful discussions. JC is partially supported by The Flatiron Institute, a division of the Simons Foundation. RQ was funded by the Deutsche Forschungsgemeinschaft (DFG, German Research Foundation) – Projektnummer 277101999 – TRR 183 (project B03), the Israel Science Foundation, and the European Research Council (Project LEGOTOP).

\bibliography{Partials.bib}

\clearpage

\onecolumngrid

\section{Supplementary Information for \\ ``Partial lattice defects in higher order topological insulators"}

\subsection{A. Look-up table for Eq. (2) in the main text and generalization to multiple types of partial dislocations.}

\begin{table}[h!]
    \centering
    \begin{tabular}{c|c|c}
          $n$  & $n\bb\cdot\bM_\nu$ & Result \\
         \hline
         even & 0 & The stacking fault characterized by $\bb$ is { consistent with} a topological insulator${}^\dag$.\\
         even & $\pi$ & The stacking fault characterized by $\bb$ is gapless.\\
         odd & 0 & The stacking fault characterized by $\bb$ is trivial.\\
         odd & $\pi$ & The stacking fault characterized by $\bb$ is a topological insulator.\\ 
    \end{tabular}
    \caption{Summary of the conclusions drawn from Eq. (2) in the main text. ${}^\dag$In this case the fault \emph{always} is a topological insulator if the fault can be made to disappear by promoting $\bb$ to a full translation and $\bb\cdot\bM_\nu=\pi$.  }
    \label{tab:partials}
\end{table}

In the main text, we considered a partial dislocation described by the Burgers vector $\bb$ for which there existed an integer $n$ such that $n\bb$ is a lattice vector; consequently, $n$ copies of the partial dislocation would be a full dislocation, for which Eq. (1) could determine the parity of helical modes.
This is applicable to atoms at high-symmetry points: for example, in a lattice with inversion symmetry, atoms at $(0,0,0)$ and $(0,0,\frac{1}{2})$ are at high-symmetry positions and it is not possible to move them away from these positions without breaking symmetry.
Table~\ref{tab:partials} provides a ``look-up table'' to summarize the possibilities.

We now consider the more general situation of a lattice with several possible Burgers vectors, denoted $\bb_i$, for which there exist integers $n_i$ such that:
\begin{equation}
    \sum_i n_i \bb_i = \bB,
    \label{eq:sumpartials}
\end{equation}
for some lattice vector $\bB$.

Now consider a crystal with $n_i$ partial dislocations of $\bb_i$ that is gapped everywhere except possibly along the dislocation cores (edges of each stacking fault), where $h_i$ helical modes reside on the dislocation characterized by $\bb_i$.
Since dislocations are additive, it follows that a full dislocation of $\bB$ must have $\sum_i n_i h_i$ helical modes.
Application of Eq.~(1) in the main text then implies:
\begin{equation}
    \sum_i n_i \bb_i \cdot \bM_\nu = \bB \cdot \bM_\nu = \sum_i n_i h_i \pi \mod 2\pi
    \label{eq:sumgeneral}
\end{equation}
Since Eq.~(\ref{eq:sumgeneral}) is defined mod $2\pi$, it can be rewritten as:
\begin{equation}
    \bB \cdot \bM_\nu = \sum_{i | n_i \text{ odd} } h_i \pi \mod 2\pi
    \label{eq:sumgeneral2}
\end{equation}
We can then conclude
\begin{enumerate}
    \item If $\bB \cdot \bM_\nu = 0$, then it is consistent that an even number of the $\bb_i$ that have odd $n_i$ may characterize dislocations that host a gapless helical mode. But it is also possible that none do. The $\bb_i$ that have even $n_i$ may or may not have a gapless helical mode.
    \item If $\bB \cdot \bM_\nu = \pi$, then there must be at least one $\bb_i$ with an odd $n_i$ that characterizes a dislocation with a gapless helical mode (and, more generally, there must be an odd number of $\bb_i$ with odd $n_i$ with a gapless helical mode.) Again, the $\bb_i$ with even $n_i$ may or may not have gapless helical modes.
\end{enumerate}
The conclusions hold as long as all of the stacking faults are gapped.

There can be additional constraints if the Hamiltonian can be deformed to a weak TI by changing the lattice symmetry without closing the bulk gap.
For example, suppose that $\bb_1 + \bb_2 = \bB$ and the Hamiltonian is deformable to a weak TI by adjusting parameters such that $\bb_1 = \bb_2$.
Then it must also be that $h_1 = h_2$ and Eq.~(\ref{eq:sumgeneral2}) simplifies to $\bB \cdot \bM_\nu = 2h_1\pi$.
Thus, if $\bB \cdot \bM_\nu = \pi$, there is a contradiction: consequently, the resulting stacking fault must be gapless!
This corresponds to the second row in Table~\ref{tab:partials}.
On the other hand, if $\bB\cdot \bM_\nu = 0$, then either $\bb_1$ and $\bb_2$ both host a helical mode, or neither one does.
This logic can be extended to however many $\bb_i$ become equal upon deforming the Hamiltonian to the weak TI phase.

\subsection{B. Correspondence between edge, screw and hinge modes in HOTIs.}\label{app:moving_fault}

\begin{figure}[tb]
    \centering
    \includegraphics[width=0.5\columnwidth]{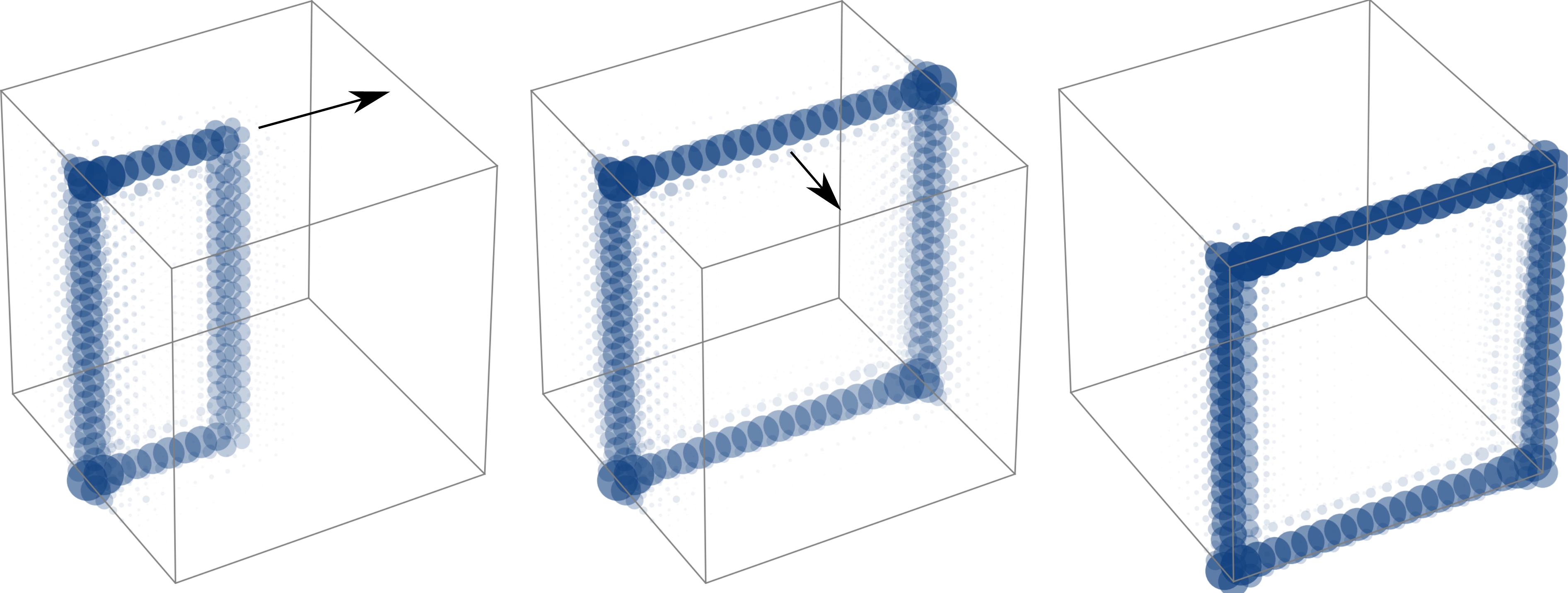}
    \caption{Correspondence between hinge and dislocation modes. Real space probability density averaged over the two states closest to $E=0$ in the extrinsic HOTI model of Eqs.~(3-MT) and (4-MT), using a finite system consisting of $20^3$ sites. Unlike Fig.~2-MT, no hinge modes appear on the top and bottom surfaces due to the opposite sign of the dimerization term, $\delta$. As shown by the arrows, it is possible to deform the stacking fault from its original position (left panel) until it covers the entire width of the system (middle panel), and afterwards move it to the system surface (right panel). The resulting system is an extrinsic HOTI, since it hosts helical modes on its hinges. As in Fig.~2-MT, larger circles and darker colors represent larger probability densities.}
    \label{fig:moving_fault}
\end{figure}

In this section, we numerically demonstrate the connection between systems hosting protected modes at partial dislocations and HOTIs. To this end, we consider the extrinsic HOTI model of Eqs.~(3-MT) and (4-MT) (main text) in the presence of a partial screw dislocation. We set $M=2$, $A=B=1$, and $t=0.3$, as in the main text, but choose an opposite sign of the dimerization, $\delta=-0.2$, such that no hinge modes appear on the top and bottom surfaces of the system (see Fig.~\ref{fig:moving_fault}). The helical mode associated to the partial dislocation can be moved by deforming the stacking fault. In this way, it is possible to obtain a system in which the stacking fault is positioned at the system surface (Fig.~\ref{fig:moving_fault}, right panel), as sketched in Fig.~1c-MT. The resulting system is an extrinsic HOTI, hosting helical modes on its hinges {due to the fact that one of the surfaces (the one containing the stacking fault) is a 2D TI} \cite{Geier2018}.

\subsection{C. Topology and low energy theory of the stacking fault.} \label{sec:lowEH}

In this section we calculate the low-energy theory of the stacking fault in both the intrinsic and extrinsic models of higher-order topological insulators presented in the main text. We follow the approach of Ref.~\onlinecite{Ran2009} by first cutting the system into two halves and then ``gluing'' them back together, but with a stacking fault. We find that the low-energy theory of the stacking fault is exactly a $\mathbb{Z}_2$ topological insulator.

We will implement the stacking fault in the $\hat{y}$-normal plane.
We start by slicing the system across that plane and computing the effective Hamiltonian on the $\hat{y}$-normal surfaces.
We begin with the intrinsic theory in Eq.~(5-MT).
To obtain the $\hat{y}$-normal surface Hamiltonian, the mass term must change sign as we cross the surface, so that $M\rightarrow M(y)$, where $0<M<4B$ in the topological regime and $M<0$ in the trivial regime.
The surface Hamiltonian is obtained by expanding around the origin (notice that at any other TRIM, $M(\bk)$ does not change sign as $M$ changes sign, so we need only consider the expansion for small $\bk$)
and replacing $k_y\rightarrow -i\partial_y$:
\begin{equation}
    H_{\rm int}(\bk \approx \mathbf{0}) =  A(k_x \sigma_z \tau_x + k_z \sigma_x \tau_x ) + \frac{m}{2}k_z\sigma_y \tau_x \mu_x + \delta\left( \sigma_x \tau_z + \sigma_y \tau_0 \right) \mu_y + \left[ M(y)\tau_z  +iA \partial_y \tau_y\right].
    \label{eq:Hint}
\end{equation}
All of the $y$-dependence is contained in the last term in brackets, which has zero energy eigenstates, exponentially localized at $y=0$, of the usual form: $\psi = e^{-\int M(y) /A \, dy} \chi$, where $\tau_x \chi = + \chi$. (This ansatz applies to the $y>0$ half of the system to be topological, where ${\rm sign}(M(y)) = {\rm sign }(y)$ and $M(y)$ approaches a constant value away from the domain wall \cite{JackiwRebbi,Khalaf2017,Queiroz2018}.)
Since the remainder of the Hamiltonian is gapped, the low-energy effective Hamiltonian is obtained by projecting onto the positive eigenvalue sector of $\tau_x$.
Denoting the projector onto this sector by $P_+$, we obtain:
\begin{equation}
    H_{\rm int}^{\rm eff}\Big|_{y=0+} = P_+[A(k_x \sigma_z \tau_x + k_z \sigma_x \tau_x ) + \frac{m}{2}k_z\sigma_y \tau_x \mu_x + \delta\left( \sigma_x \tau_z + \sigma_y \tau_0 \right) \mu_y]P_+= A(k_x \sigma_z + k_z \sigma_x) + \frac{m}{2}k_z \sigma_y \mu_x + \delta \sigma_y \mu_y 
    \label{eq:HeffR}
\end{equation}
The same ansatz applies to the other ($y<0$) half of the system with $M(y) \rightarrow -M(y)$; denoting the projector onto the negative eigenvalue sector of $\tau_x$ by $P_-$, we obtain:
\begin{equation}
    H_{\rm int}^{\rm eff}\Big|_{y=0^-}=P_- [A(k_x \sigma_z \tau_x + k_z \sigma_x \tau_x ) + \frac{m}{2}k_z\sigma_y \tau_x \mu_x + \delta\left( \sigma_x \tau_z + \sigma_y \tau_0 \right) \mu_y]P_- = -A(k_x \sigma_z + k_z \sigma_x) - \frac{m}{2}k_z \sigma_y \mu_x + \delta \sigma_y \mu_y 
    \label{eq:HeffL}
\end{equation}
The combined Hamiltonian of the 2D interface without a defect is given by:
\begin{equation}
    H_{\rm int}^\text{no fault} \equiv H_{\rm int}^{\rm eff}\Big|_{y=0+} +  H_{\rm int}^{\rm eff}\Big|_{y=0^-} = 
    \left( A(k_x \sigma_z + k_z \sigma_x) + \frac{m}{2}k_z \sigma_y \mu_x\right)\tau_x + \delta \sigma_y \mu_y \tau_0
    \label{eq:Hnofault}
\end{equation}

Now we must implement the domain wall. We do this by translating half the system by half a lattice translation in the $\hat{z}$ direction; since this translation exchanges the sublattices, it is implemented by the $\mu_x$ operator.
Noticing that only the $\delta$ term is odd under $\mu_x$, the domain wall is implemented by taking, for example, $\delta<0$ for $y=0^-$ in Eq.~(\ref{eq:HeffL}) and $\delta>0$ for $y=0^+$ in Eq.~(\ref{eq:HeffR}).
Then, the low-energy theory of the stacking fault is given by:
\begin{equation}
    H_{\rm int}^{\rm fault} \equiv H_{\rm int}^{\rm eff}\Big|_{y=0+} + \mu_x \left( H_{\rm int}^{\rm eff}\Big|_{y=0^-} \right)\mu_x = 
    \left( A(k_x \sigma_z + k_z \sigma_x) + \frac{m}{2}k_z \sigma_y \mu_x\right) \tau_x + \delta \sigma_y \mu_y  \tau_x
    \label{eq:Hfault}
\end{equation}
Comparing Eqs.~(\ref{eq:Hnofault}) and (\ref{eq:Hfault}) shows that they are 2D insulators which differ by a single band inversion at the origin.
Thus, one of the insulators is a 2D TI and the other is a 2D normal insulator.
Since a 2D TI always has a 1D helical modes at its edge, there must be a 1D helical mode at the edge of the stacking fault.
This also proves that the stacking fault realizes an ``embedded'' 2D TI, as introduced in Ref.~\onlinecite{Tuegel2018}.

We can repeat this procedure for the extrinsic Hamiltonian.
Expanding around the origin and replacing $k_y\rightarrow -i\partial_y$ yields,
\begin{equation}
    H_{\rm ext}(\bk \approx \mathbf{0}) =- Ak_x \sigma_x \tau_x + \sigma_z\tau_x (t k_z\mu_x  -2\delta\mu_y + \delta k_z \mu_x) + \left[  M(y)\tau_z  - i \partial_y A \sigma_y \tau_x \right]
    \label{eq:Hext}
\end{equation}
The zero-energy eigenstates of the term in brackets are again of the form $e^{-\int M(y)/A \, dy} \chi$, where now $\chi$ satisfies $\sigma_y\tau_y\chi = + \chi$. 
Since there are two states with positive eigenvalue of $\sigma_y\tau_y$, we use $\tilde{\sigma}_{x,y,z}$ to denote a new set of Pauli matrices that acts on these two states.
Then the projection of Eq~(\ref{eq:Hext}) onto the low-energy sector yields:
\begin{equation}
    H_{\rm ext}^{\rm eff} \Big|_{y=0^+}  = Ak_x\tilde{\sigma}_z -k_z\tilde{\sigma}_x(t+\delta)\mu_x + 2\delta \tilde{\sigma}_x \mu_y.
    \label{eq:HeffR2}
\end{equation}
Similarly,
\begin{equation}
    H_{\rm ext}^{\rm eff} \Big|_{y=0^-}  = -Ak_x\tilde{\sigma}_z -k_z\tilde{\sigma}_x(t+\delta)\mu_x + 2\delta \tilde{\sigma}_x \mu_y,
    \label{eq:HeffL2}
\end{equation}
which is identical to Eq.~(\ref{eq:HeffR2}) after a basis change of conjugating by $\tilde\sigma_x$.
In the basis used for the extrinsic Hamiltonian in Eqs.~(3-MT) and (4-MT), translation by $\hat{z}/2$ is implemented by the matrix $e^{ik_z/2}(\cos \frac{k_z}{2} \mu_x + \sin \frac{k_z}{2} \mu_y)$. The terms proportional to $\delta$ are odd under this partial translation, while all other terms are even (this is also true exactly in Eqs.~(5-MT) and (6-MT) and to leading order in $k_z$ in Eqs.~(\ref{eq:HeffR2}) and (\ref{eq:HeffL2}).)
Thus, as in the intrinsic case, we implement the domain wall by taking, for example, $\delta<0$ for $y=0^-$ in Eq.~(\ref{eq:HeffL2}) and $\delta>0$ for $y=0^+$ in Eq.~(\ref{eq:HeffR2}).
Then, exactly as in the intrinsic case, the low-energy theory of the stacking fault differs by the interface without a defect by exactly one band inversion.
Thus, there must be a helical mode at their interface.

Note that there is one more subtlety compared to the intrinsic case, which is that when $M$ changes sign, the full mass term, $M(\bk)$, changes sign not only at $k_x = k_y = k_z = 0$, but also at $k_x = k_y = 0, k_z =\pi$ (since $M(\bk)$ is independent of $k_z$.)
Thus, we must also consider the low-energy theory expanded around $k_x = k_y=0, k_z=\pi$.
The effect in Eqs.~(\ref{eq:HeffR2}) and (\ref{eq:HeffL2}) is to swap $t\leftrightarrow -\delta$.
When we implement the stacking fault by taking $\delta\rightarrow -\delta$ in only Eq.~(\ref{eq:HeffL2}) (but not Eq.~(\ref{eq:HeffR2})), we do not change the sign of the mass term; that is, both Eqs.~(\ref{eq:HeffR2}) and (\ref{eq:HeffL2}) have the same mass, whether or not there is a stacking fault.
Hence, the point $k_x=k_y=0, k_z=\pi$ does not contribute an extra band inversion to the stacking fault and so does not affect our conclusion that a helical mode must exist at the edge of the stacking fault.

\subsection{D. Bandstructures associated to lattice defects.}\label{app:bands}

In this section, we provide additional numerical details on the partial screw dislocations in both extrinsic and intrisic HOTIs, as well as the full screw dislocation in a WTI.
For the intrinsic and extrinsic HOTI Hamiltonians, we use the same parameters as in the main text. To study a WTI Hamiltonian, we start from the extrinsic model of Eqs.~(3-MT) and (4-MT) and set the dimerization to $\delta=0$, such that QSH layers are equally coupled.

\begin{figure}[b]
    \centering
    \includegraphics[width=0.75\columnwidth]{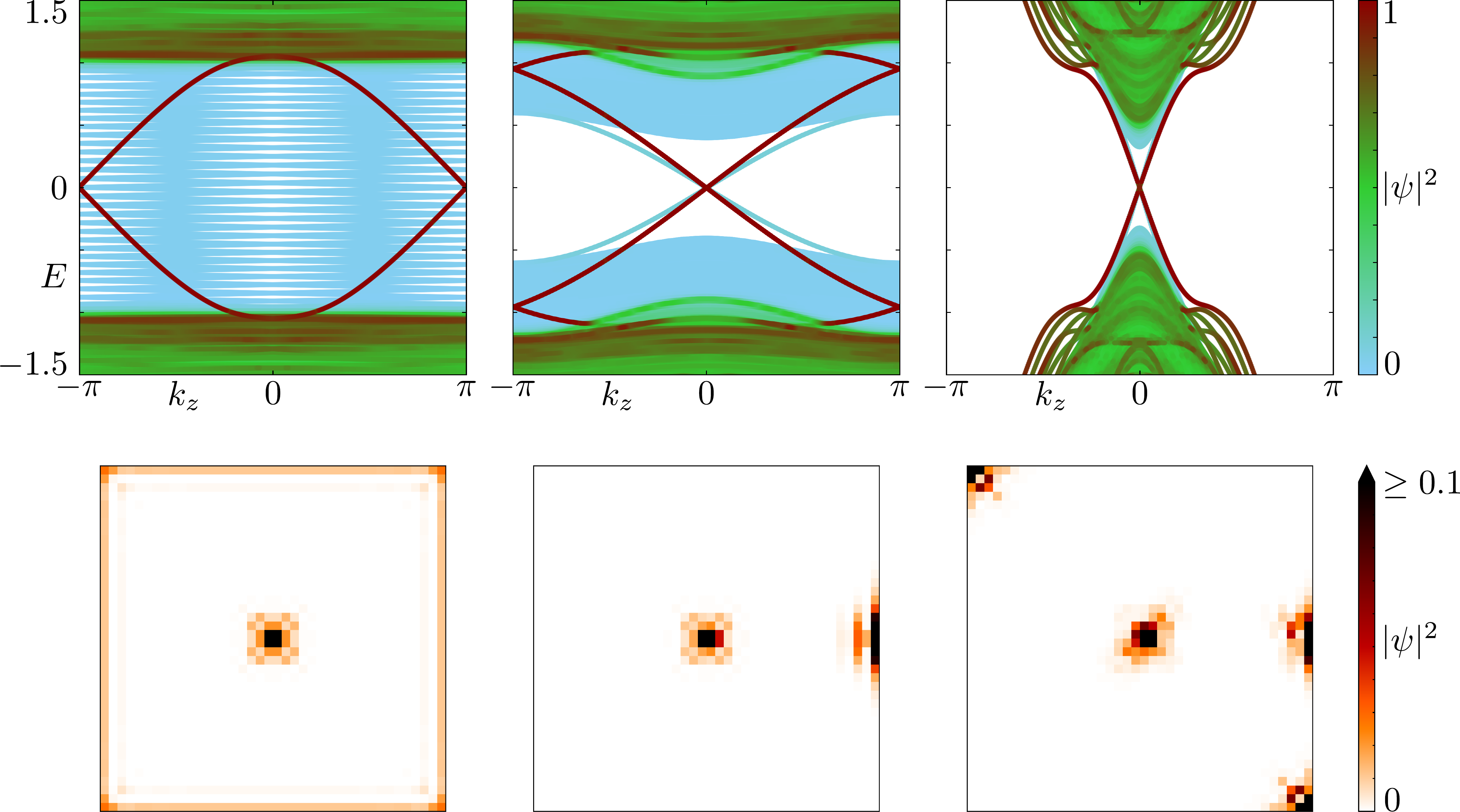}
    \caption{Bandstructures (top) and real space probability densities (bottom) associated to a WTI (left), an extrinsic HOTI (middle), and an intrinsic HOTI (right). Results are obtained in an infinite column geometry, infinite along the $z$ direction and with a cross section of $20\times20$ unit cells (top) or $40\times40$ unit cells (bottom) in $x$ and $y$. In the top panels, the color scale represents the integrated probability density over half of the system sites which are closest to the center of the system. As such, states localized at the core of the screw dislocation are shown in red, states close to the column boundary (surfaces and hinges) are colored blue, whereas bulk states are green. The bottom panels show the probability densities of $E=0$ states for $k_z=\pi$ (left) and $k_z=0$ (middle and right).}
    \label{fig:bands}
\end{figure}

Figure \ref{fig:bands} shows band structures and real space probability densities obtained in an infinite column geometry: the system is infinite in the $z$ direction and finite with a square cross-section in $x$ and $y$. For the WTI (Fig.~\ref{fig:bands} left), we find an $E=0$ doubly degenerate state at $k_z=\pi$, which is localized at the core of the (full) screw dislocation. This state is separated in real space from the gapless surfaces of the WTI. Turning on a nonzero dimerization term $\delta=0.2$ doubles the unit cell and opens a gap in the WTI surface (Fig.~\ref{fig:bands} middle). The helical mode at the center of the column survives, remaining localized at the core of the (now partial) screw dislocation. Due to the folding of the Brillouin zone, the state now appears at $k_z=0$. A second $E=0$ doubly degenerate state exists on the system surface, marking the point where the stacking fault intersects the system boundary. Finally, in the intrinsic HOTI model we find a total of four helical modes at $E=0$, positioned at $k_z=0$  (Fig.~\ref{fig:bands} right). The modes overlap in energy and momentum, such that only the screw mode is visible in the bandstructure (red lines). The real space positions of these modes are the boundaries of the stacking fault (on the partial screw dislocation and on the system surface), as well as on two of the hinges.

For all three systems, the zero energy in gap states occur at time-reversal invariant momenta, are doubly degenerate due to Kramers' theorem, and are separated from each other in real space, as shown in the bottom panels of Fig.~\ref{fig:bands}. Therefore, they cannot be removed by any infinitesimal, time-reversal symmetric perturbation, and are consequently topologically protected, as discussed in the previous section.

\subsection{E. Helical modes at partial edge dislocations.}\label{app:partial_edge}

\begin{figure}[tb]
    \centering
    \includegraphics[width=0.5\columnwidth]{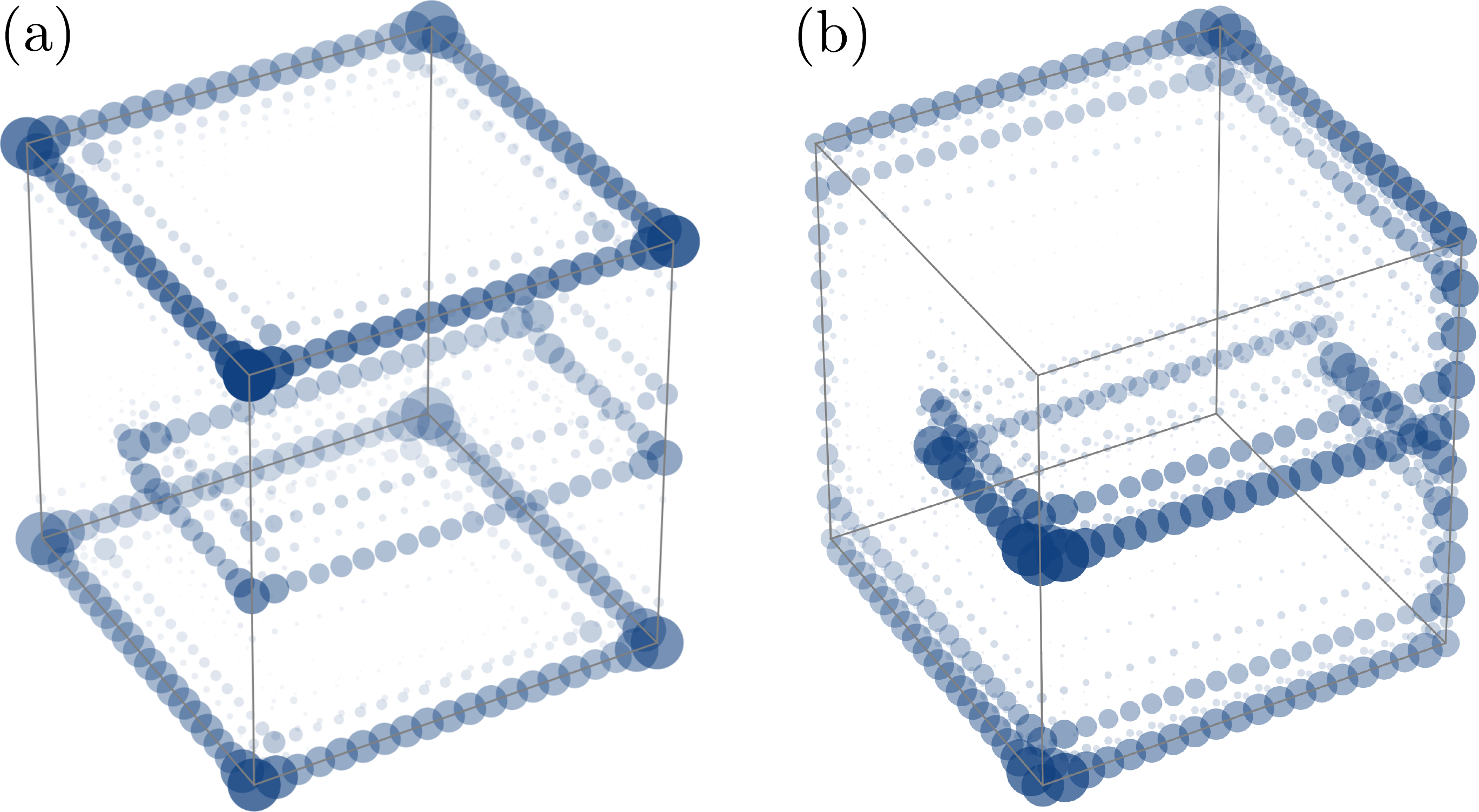}
    \caption{Partial edge dislocations. Real space probability density averaged over the \textcolor{red}{}ten (left) and eight (right)} states closest to $E=0$, in the extrinsic (panel a) and intrinsic (panel b) HOTI models. Both systems contain a partial edge dislocation, and in both cases helical modes appear at the boundaries of the stacking fault. Model parameters and circle size/color are the same as in Fig.~2-MT.
    \label{fig:edge_dis}
\end{figure}

As discussed in the main text and sketched in Fig.~1, helical modes can appear not only at partial screw, but also at partial edge dislocations. We demonstrate this numerically in Fig.~\ref{fig:edge_dis}, both for the extrinsic and for the intrinsic HOTI models. In both cases, we use the same Hamiltonian parameters {and system sizes} as in the main text. The edge dislocations are produced by removing a half-plane of sites at constant $z$ coordinate, corresponding to the $a$ sublattice. In the intrinsic HOTI model, sites on the top and bottom of the stacking fault are still connected by hopping terms within the same sublattice (from $b$ to $b$). In the extrinsic model, removing half of an $a$ sublattice layer produces sites which are decoupled across the stacking fault, since the Hamiltonian only contains nearest neighbor hopping terms in the $z$ direction. We reconnect these sites using a hopping term with the same matrix structure as the rest of the bulk, $t \sigma_z\tau_x$.

\bibliography{Partials.bib}

\end{document}